\documentclass[twocolumns,showpacs,amsmath,amssymb]{revtex4}
\usepackage{graphicx}
\usepackage{epsfig}
\usepackage{graphicx}% Include figure files
\usepackage{dcolumn}% Align table columns on decimal point
\usepackage{bm}% bold math
\begin{document}
\title{ Phantom Energy with Variable $G$ and $\Lambda$}
\author{Arbab I. Arbab}
\email{aiarbab@uofk.edu} \affiliation{Department of Physics,
Faculty of Science, University of Khartoum, P.O. Box 321, Khartoum
11115, Sudan}\affiliation{Department of Physics and Applied
Mathematics, Faculty of Applied Sciences and Computer, Omdurman
Ahlia University, P.O. Box 786, Omdurman, Sudan}
\date{\today}
\begin{abstract}
We have investigated a cosmological model of a phantom energy with
a variable cosmological constant ($\Lambda$) depending on the
energy density ($\rho$) as $\Lambda\propto \rho^{-\alpha}$,
$\alpha=\rm const.$ and a variable gravitational constant ($G$).
The model requires $\alpha<0$ and a negative gravitational
constant. A negative gravitational constant may forbid \emph{black
holes} to form a particle horizon in a background of phantom
energy. This implies that black holes are naked, and consequently
the \emph{Cosmic Censorship} theorem is violated. The cosmological
constant evolves with time as, $\Lambda\propto t^{-2}$. For
$\omega>-1$ and $\alpha<-1$ the cosmological constant,
$\Lambda<0$,  $G>0$ and $\rho$ decrease with cosmic expansion. For
ordinary matter (or dark matter), i.e., $\omega>-1$ we have
$-1<\alpha<0$ and $\beta>0$ so that  $G>0$ increases with time and
$\rho$ decreases with time. Cosmic acceleration with dust
particles is granted provided $-\frac{2}{3}<\alpha <0$ and
$\Lambda>0$.
\end{abstract}
\pacs{98.80.-k, 98.80.Es, 98.80.Hw}
%\keywords{Phantom energy, dark
%energy, cosmological constant, cosmic acceleration}
\preprint{APS/123-QED}
 \maketitle
\section{Introduction}
Cosmologists have wondered  whether our present universe will
eventually re-collapse and end with a Big Crunch, or expand
indefinitely and eventually becomes cold and empty. However,
recent evidence from supernovae type I ushers into  a flat
universe, possibly with a cosmological constant or some other sort
of negative-pressure dark energy, has suggested that our fate is
accelerating (Perlmutter \emph{et al.}, Riess \emph{et al.}).
However, the data may actually be pointing toward an astonishingly
different cosmic end game. Caldwell \emph{et al.} explored the
consequences that follow if the dark energy is a phantom energy,
i.e., the sum of the pressure and energy density is negative. The
positive phantom-energy density becomes infinite in a finite time,
overcoming all other forms of matter, that will rapidly brings the
epoch of cosmic structure to a halt (Dabrowski \emph{at al.}). The
phantom energy rips apart every bound matter before the Universe
ends into a {\tt Big-Rip}.

However, the phantom energy scenario does violate the \emph{the
strong energy condition} (SEC), a principle that keeps energies
positive and imposes energy conservation on a global scale. It is
the strong energy condition that helps to rule out  wormholes,
warp drives, and time machines. Dark energy requires an equation
of state $p+3\rho<0$. The violation of the null energy condition
(NEC) $p+\rho<0$ results in energy flows  faster than the speed of
light. A phantom behavior is predicted by several scenarios, e.g.,
kinetically driven models (Chiba \emph{et al.}) and some versions
of braneworld cosmologies ( Sahni and  Shtanov).

Another possibility for dark energy is an energy of a scalar field
known as \emph{quintessence} having an equation of state such that
$-1<\omega<-\frac{1}{3}$ (Caldwell \emph{et al.}, Ratra \&
Peebles, Wetterich, Turner \& White, Caldwell). Assuming the
quintessence field coupled minimally to gravity, one writes it
lagrangian as,
\begin{equation}
{\cal L}=\frac{1}{2}\dot\phi^2+V(\phi)\ ,
\end{equation}
with energy density and pressure given by
\begin{equation}
\rho=\frac{1}{2}\dot\phi^2+V(\phi)\ ,\qquad
p=\frac{1}{2}\dot\phi^2-V(\phi).
\end{equation}
 where a dot is a
derivative \emph{w.r.t} time. In the so-called 'tracker' models
the scalar field density (and its equation of state) remains close
to that of the dominant background matter during most of
cosmological evolution. The equation of state is given by
\begin{equation}
\frac{p}{\rho}=\frac{\frac{1}{2}\dot\phi^2-V(\phi)}{\frac{1}{2}\dot\phi^2+V(\phi)},
\end{equation}
  A comprehensive study of quintessence is
investigated by Ratra and Peebles. However, A tracker potential of
the form $V(\phi)\propto \phi^{-n}$, $n=\rm const.$ is considered
recently  by Sahni. \\
A minimally coupled scalar field to gravity
has the general lagrangian of the form
$${\cal L}=P\equiv P(X,\phi),$$
where $X=\frac{1}{2} g_{\mu\nu}\partial^\mu\phi\partial^\nu\phi$,
having an energy momentum tensor
$$ T_{\mu\nu}=-{\cal L}g_{\mu\nu}+P'\partial_\mu\phi\partial_\nu\phi,$$
or
$$T_{\mu\nu}=P'\partial_\mu\phi\partial_\nu\phi-Pg_{\mu\nu}.$$
If X is time-like vector then $T_{\mu\nu}$ is equivalent to that of
 a perfect fluid,
$T_{\mu\nu}=(p+\rho)u_\mu u_\nu-pg_{\mu\nu}$ with energy density
$\rho=2XP'-P$, pressure $P$  and 4-velocity
$u_\mu=\frac{\partial_\mu \phi}{\sqrt{2X}}$. Its equation of state
amounts to
$$\omega=\frac{P}{\rho}=-1+\frac{2XP'}{\rho}=-1+\frac{P'\dot\phi^2}{\rho},$$
where $P'=\frac{\partial P}{\partial X}$. Since for dark matter
$T_{00}=\rho>0$, one has the condition $P'<0$ for $\omega<-1$. The
current observational data amounts to $\omega<-1$, however this
poised a problem.

 Moreover, one may include the possibility of an equation of state $p=-\rho$. This
is attributed to existence of vacuum energy or the cosmological
constant. At the present time it is difficult to tell which form of
energy our universe consists of.

In this paper, we investigate the evolution of dark energy and
phantom energy arising from the introduction of a cosmological
constant that evolves as $\Lambda=\frac{3\beta}{\rho^\alpha}$,
where $\alpha,\ \beta=\rm const.$. With this assumption, a phantom
energy arises whenever $p+\rho<0$ and $\alpha>0$. However,  the
gravitational constant becomes negative. In the present model, the
dark energy models do no necessarily require the condition
$p>-\frac{1}{3}\rho$. Cosmic acceleration is generated for
$\alpha>0$, $-1<\alpha <0$ and $\alpha<-1$. Phantom energy with
variable $G$ has been recently considered by Stefancic. We have
shown that during the evolution of the domain-walls the
cosmological constant flips it sign. We have seen that the whole
evolution of the universe is characterized by the the two
constants $\alpha$ and $\beta$.
\section{The Model}
Consider the Einstein-Hilbert action with a cosmological constant
term ($\Lambda$)
\begin{equation}
S=-\frac{1}{16\pi G}\int d^4x\sqrt{g}(R+2\Lambda)+S_{\rm matter}
\end{equation}
The variation of the metric with respect to $g_{\mu\nu}$ with
$f(R)=R-2\Lambda$,  gives (Amarzguioui \emph{et al.})
\begin{equation}
f'(R)R_{\mu\nu}-\frac{1}{2}f(R)g_{\mu\nu}=-8\pi G T_{\mu\nu}\ ,
\end{equation}
where $T_{\mu\nu}$ is the energy momentum tensor of the cosmic
fluid.
\\
For an ideal fluid one has
\begin{equation}
T_{\mu\nu}=(\rho+p)u_\mu u_\nu+pg_{\mu\nu},
\end{equation}
where $u_\mu, \rho, p$ are the velocity, density and pressure of the
cosmic fluid. Contracting Eq.(5), using Eq.(6) and taking the 00
components give, the equation
\begin{equation}
 Rf'(R)-2f(R)+8\pi GT=0,
\end{equation}
and
\begin{equation}\label{2}
f'(R)R_{00}+\frac{1}{2}f(R)+8\pi G\ T_{00}=0,
\end{equation}
with $T_{00}=\rho$, $T=\rho-3p$ and $T_{ij}=-p$ for $i,j=1,2,3$. For
a flat Friedmann-Lemaître-Robertson-Walker metric,
$$ds^2=dt^2-a^2(t)\left(dr^2+r^2(d\theta^2+\sin^2\theta\ d\phi^2)\right),$$
one has $R_{00}=-3\frac{\ddot a}{a}$ and $R=-6[(\frac{\dot
a}{a})^2+\frac{\ddot a}{a}]$, so that Eqs.(7) and (8) yield
\begin{equation}\label{1}
 3\left(  \frac{\dot a}{a}\right)^2=8\pi \ G\rho +\Lambda,
\end{equation}
\begin{equation}\label{1}
3 \left(  \frac{\ddot a}{a}\right)=-4\pi\ G(\rho+3p) +\Lambda,
\end{equation}
and the energy conservation equation reads,
\begin{equation}\label{1}
   \dot\rho+3\left(\frac{\dot a}{a}\right)(\rho+p)=0.
\end{equation}
The pressure $p$ and energy density $\rho$ of an ideal fluid are
related by the equation of state,
\begin{equation}
p=\omega\ \rho,\qquad\omega=\rm const.
\end{equation}
The Einstein field equation, with time-dependent $G$ and $\Lambda$,
then yields two independent equations (Eqs.(9) and (10)) having the
same form as in the standard model. Hence, we now allow $\Lambda$
and $G$ to vary with time, i.e., $\Lambda=\Lambda(t)$ and
$G=G(t)$.\\
The Bianchi identity
\begin{equation}\label{1}
(R^{\mu\nu}-\frac{1}{2}Rg^{\mu\nu})_{;\ \mu}=-(8\pi
GT^{\mu\nu}+\Lambda g^{\mu\nu})_{;\ \mu}=0,
\end{equation}
and Eqs.(6) and (12) imply that
\begin{equation}\label{1}\nonumber
 G\dot\rho+3(1+\omega)\rho G\frac{\dot a}{a}+\rho\dot
G+\frac{\dot\Lambda}{8\pi}=0\ ,
\end{equation}
 so that the energy conservation, Eq.(11), entitles that (Beesham,
Abdel Rahman, Arbab)
\begin{equation}\label{1}
8\pi \dot G\rho+\dot\Lambda=0.
\end{equation}
We consider here the ansatz
\begin{equation}\label{1}
   \Lambda=\frac{3\beta }{\rho^\alpha} \ ,\qquad \beta\ , \ \alpha=\rm const.
\end{equation}
Integrating Eq.(11), using Eq.(12), we obtain
\begin{equation}\label{0}
    \rho=A a^{-3(1+\omega)}\ , \ A=\rm const.
\end{equation}
Substituting this in Eq.(13) using Eq.(14),  one gets
\begin{equation}\label{2}
G=\left(\frac{-3\alpha\beta}{8\pi(1+\alpha)A^{1+\alpha}}\right)a^{3(1+\omega)(1+\alpha)},
 \qquad \alpha\ne-1.
\end{equation}
When $\alpha=0$, Eq.(13) implies that $\Lambda=\rm const.$ and
$G=\rm const.$. Substituting Eqs.(15) and (16) into Eq.(9) we
obtain, $H=\frac{\dot a}{a}$,
\begin{equation}\label{5}
    H^2=\left(\frac{\beta}{1+\alpha}\right)A^{-\alpha}a^{3(1+\omega)\alpha},  \qquad \alpha\ne-1.
\end{equation}
or
\begin{equation}\label{5}
    H^2=\left(\frac{\beta}{1+\alpha}\right)\rho^{-\alpha}\ , \qquad \alpha\ne-1.
\end{equation}
Using Eq.(18), Eq.(16) can be written as
\begin{equation}\label{5}
    8\pi G=\left(\frac{-3\alpha\beta}{1+\alpha}\right)\rho^{-(\alpha+1)}\ , \qquad \alpha\ne-1.
\end{equation}
Integrating Eq.(18), one obtains,
\begin{equation}\label{5}
    a(t)=\left(\frac{F}{A}\right)^{-1/3(1+\omega)}\ t^{-2/3\alpha(1+\omega)},\qquad
    \alpha\ne
    -1,
\end{equation}
where
$F=\left[-\frac{3}{2}\alpha(1+\omega)(\frac{\beta}{(1+\alpha)})^{1/2}\right]^{2/\alpha}$
Using Eq.(21), Eq.(17) becomes
\begin{equation}\label{2}
G=\frac{-3\alpha\beta}{8\pi\ (1+\alpha)}F^{-(1+\alpha)}\
t^{-2(1+\alpha)/\alpha},\qquad \alpha\ne-1.
\end{equation}
Using Eq.(21), Eq.(16) reads
\begin{equation}\label{2}
\rho=F\ t^{2/\alpha},
 \qquad \alpha\ne0, \ -1,
\end{equation}
which shows that the energy density of the phantom increases with
expansion, since $\alpha>0$. Using Eq.(23), Eq.(15) yields
\begin{equation}\label{2}
\Lambda=\frac{3\beta}{F^\alpha}t^{-2}.
\end{equation}
This form is found to emerge from many of the vacuum decaying
models.
\section{Phantom energy}
For an expanding universe, we require that $-\alpha(1+\omega)>0$.
Since, $\alpha>0$, one must have the relation $\omega+1<0$. This is
the familiar condition for the existence of phantom energy. This
relation implies that $p<-\rho$. In this case, the energy density
grows with time, as it is  evident from Eq.(23). This is the
condition for phantom energy. It is very interesting to see that
such a cosmological constant variation leads to phantom energy
solution. We, however, notice that phantom energy existence requires
$G<0$, so that $\dot G>0$ and $\dot\Lambda<0$. We, therefore, see
that the phantom energy has negative gravity. One may attribute this
to the anti-gravitating nature of phantom energy. Such a bizarre
behavior could be the reason why  phantom energy has a negative
pressure, unlike the ordinary matter.

In the scalar field theory, the phantom energy is model by a field
with a negative kinetic energy, but negative pressure and positive
energy. In our present scenario, $G<0$ and $p<0$, but $\rho>0$.
Hence, the two pictures of phantom energy evolution could be
equivalent. However, in our present scenario both the
gravitational and the cosmological constants decrease with time.
Unlike the standard phantom energy model, we see that as
$t\rightarrow 0$ $a\rightarrow 0$, $\rho\rightarrow 0$; and as
$t\rightarrow \infty$ $a\rightarrow \infty$, $\rho\rightarrow
\infty$. Consider now the case $\omega=-\frac{3}{2}$ and
$\alpha=\frac{2}{3}$. In this case, $a\propto t^2$, and hence,
$\rho\propto t^3$ and $G\propto t^{-5}$. An equivalent case
corresponds to $\alpha=1$ and $\omega=-\frac{4}{3}$. However, in
the latter case $\rho\propto t^2$ and $G\propto t^{-4}$. The
present observational data favor negative values for $\omega$
rather than positive ones. In particular, models with $-1.62 <
\omega < -0.74$ are favored observationally (Carroll \emph{at
al.}).
\section{Negative Gravitational Constant}
For phantom energy to develop, we require the gravitational
constant to be negative during its dominance. Phantom energy
implies that $\rho+p<0$ and this also implies $\rho+3p<0$. The
gravitational potential ($V$) for matter having energy
density($\rho$) and pressure ($p$), satisfies the Poisson equation
$$\nabla^2V=4\pi G(\rho+3p).$$
For phantom energy one has $$\rho+3p<0\qquad\text{and}\qquad G
<0$$ and
 so that the above equation  does not change sign. This would
mean that this equation governs phantom too. A \emph{black hole}
metric (Schwarzschild) with phantom energy does not lead to a
particle horizon, since the Schwarzschild  metric
$$ds^2=(1-\frac{2GM}{rc^2})\ dt^2-\frac{dr^2}{(1-\frac{2GM}{rc^2})}-r^2(d\theta^2+\sin^2\theta \ d\phi^2)$$
will be defined everywhere (except at $r=0$). Thus, a spherical
phantom star of mass $M$ would not be bounded by an event horizon.
Hence, a phantom energy prohibits a black hole to be covered by a
horizon. Hence, black holes become naked, and consequently
\emph{Cosmic Censorship} theorem is violated.  It is recently
shown by Cai1 and Wang that when only dark energy (as a background
energy with $p<-\frac{1}{3}\rho$) is present, black holes are
never formed.  The  area of the event horizon of a non-rotating
\emph{black hole}  with electric charge $Q$ and mass $M$
determined by
 Reissner–Nordström as
 $$A_H=\frac{4\pi G^2}{c^4}\left(M+\sqrt{M^2-Q^2/G}\right)^2.$$
For a negative decreasing gravitational constant, one has
approximately,
$$A_H\backsimeq\frac{4\pi (-G)Q^2}{c^4}>0.$$
Hence, a decreasing gravitational constant implies a reduction in
the area of the event horizon. This means that a \emph{black hole}
would eventually disappear during the course of cosmic expansion.
It has been shown by Babichev \emph{et al.} that \emph{black
holes} accreing phantom energy will lose mass and dissappear.
\section{Cosmic Acceleration}
The present observational data usher toward an accelerated expansion
of the universe (Reiss, \emph{et al.}, Perlmutter \emph{et al.}).
The deceleration parameter is given by $q=-\frac{\ddot a}{a}/H^2$.
Using Eq.(21), this yields
\begin{equation}
q=-\frac{2+3\alpha (1+\omega)}{2}.
\end{equation}
For an accelerating universe, one must have $q<0$, i.e., $\alpha
>\frac{-2}{3(1+\omega)}$. For non-phantom energy one has
$1+\omega>-1$ so that $\alpha<0$. We treat the case $\alpha=-1$
separately.

For $\omega>-1$ and $\alpha<-1$, the scale factor grows with time.
For a positive gravitational constant, $G>0$, Eq.(20) implies
$\beta<0$, so that the cosmological constant becomes negative,
$\Lambda <0$. Eqs.(22) and (23) imply that $G$ and $\rho$ decrease
with cosmic expansion. For $\alpha=-2$ and $\omega=-\frac{1}{2}$ one
has $a\propto t^{2/3}$. This dark energy mimics the evolution of
dust particles. Similarly for $\alpha=-2$ and $\omega=-\frac{1}{3}$,
the scale factor evolves as, $a\propto t^{1/2}$. Hence, cosmic
strings in this case evolves like radiation. For the two case the
energy density evolves as, $\rho\propto t^{-1}$ and $G\propto
t^{-1}$.
\\
Now consider the case $-1<\alpha<0$, $\beta>0$ and $\omega>-1$. This
is the case for ordinary matter (or dark energy). In this case,
$G>0, \Lambda>0$. Eqs.(22) and (23) imply that $G$ increases with
time and $\rho$ decreases with time. For $\alpha=-\frac{1}{2}$ and
$\omega=1$, the scale factor varies as, $a\propto t^{2/3}$. Thus,
stiff matter mimics ordinary matter. In this case, $G\propto t^{2}$
and $\rho\propto t^{-4}$. We notice that in the present scenario one
has the relation $G\rho\propto H^2$. Such a relation is known to
satisfy the Machian cosmology (Arbab, 1997).
\subsection{Acceleration with ordinary matter}
In this model, it is possible to have a cosmic acceleration in the
present epoch with ordinary matter (dust or radiation).  To my
knowledge this solution has not been considered before. This arises
due to the presence of the cosmological constant of the form
suggested in Eq.(15). We consider here $\omega>0$ and $\beta>0$. In
this case we have:
\begin{itemize}
    \item dust: $\omega=0\ \Rightarrow \alpha>-\frac{2}{3}.$
    \item radiation: $\omega=1\ \Rightarrow \alpha>-\frac{1}{2}.$
    \end{itemize}
For the two cases, both $G>0$ and $\Lambda>0$. Moreover, $G$
increases while $\Lambda$ decreases with cosmic time.
\subsection{Acceleration with dark energy}
In this case, we consider $-1<\omega<0$. Cosmic acceleration for
strings-like and domain-walls like fluid respectively imply
\begin{itemize}
    \item cosmic strings: $\omega=-\frac{1}{3}\ \Rightarrow \alpha>-1.$
    \item domain-walls: $\omega=-\frac{2}{3}\ \Rightarrow \alpha>-2.$
\end{itemize}
We observe that domain-walls proceeds in two different ways:\\ (i)
when $\alpha<-1$, $\beta<0$ for $G>0$  so that $\Lambda<0$.\\ (ii)
when $\alpha>-1$, $\beta>0$ for $G>0$  so that $\Lambda>0$. Hence,
during the domain-walls evolution the cosmological constant
changes its sign.
\section{New Solution}
We study in this section the case $\alpha=-1$. Eqs.(14) and (15)
imply that
\begin{equation}
G=\frac{3\beta}{8\pi}(1+\omega)\ln (Ca)\ ,\qquad C=\rm const.
\end{equation}
This equation implies that when $a\rightarrow 0$,
$G\rightarrow-\infty$ unless $\beta$ or $1+\omega$ becomes negative.
Physically this case is meaningless. Hence, one may say that in the
case $\Lambda=3\beta\rho$, the initial singularity is avoided
provided the gravitational constant is allowed to vary with time.
For an expanding universe the gravitational constant must increase
with cosmic time. We notice that $G$ increase with time for
$\beta>0$ ($\Lambda>0$) and $\omega>-1$ (ordinary/dark energy), or
$\beta<0$ ($\Lambda<0$) and $\omega<-1$ (phantom energy). Upon using
this equation, Eq.(9) becomes
\begin{equation}
\dot a^2=\beta\left[1+3(1+\omega)\ln (Ca)\right]Aa^{-(1+3\omega)}.
\end{equation}
The solution of the above equation gives the time dependence of
the gravitational constant and energy density.
\section{Inflationary Solution}
We consider  here the case $\omega=-1$. In this case Eq.(16) yields
$\rho=\rm const.$ and then  Eq.(15) implies $\Lambda=\rm const.$ Now
Eq.(14) implies that $G=\rm const$. and hence Eq.(9) gives $a\propto
\exp(H_0t)$, $H_0=\rm const.$. This is the standard de-Sitter
expansion. We notice from Eq.(26) that during inflation the
gravitational constant vanishes. This may have assisted the universe
to inflate with  a constant expansion rate,
$H=\sqrt{\frac{\Lambda}{3}}$.
\section{Concluding Remarks}
We have considered in this paper a cosmological model with a
cosmological constant varies as
$\Lambda=\frac{3\beta}{\rho^\alpha}$. We have found that cosmic
acceleration is guaranteed in radiation ($-\frac{1}{2}<\alpha <0$)
and matter ($-\frac{2}{3}<\alpha <0$) dominated epochs with
$\Lambda>0$ and $G>0$. For dark energies (cosmic-strings/domain
-walls) cosmic acceleration occurs when $\alpha<-1$. Phantom
energy with $\omega<-1$ is allowed provided $\alpha>0$ and $G<0$.
The phantom energy density varies as, $\rho\propto t^{2/\alpha}$.
For instance, for $\alpha=1$ and $\omega=-\frac{4}{3}$ the scale
factor increases as, $a\propto t^2$. The negative gravitational
constant is tantamount to negative kinetic energy for phantom
field. The case $\omega=-1$ gives the familiar de-Sitter
inflationary solution.

\section{Acknowledgments}
I would like to thank M. Jamil for useful communication.
%\newpage
\section{References}
\hspace{-0.5cm} A. Beesham, \emph{Int. J. Theor. Phys.}, 25, 1295 (1986).\\
T. Chiba, T. Okabe, and M. Yamaguchi, \emph{Phys. Rev}. D 62,
023511 (2000).\\
V. Sahni and Y. Shtanov, \emph{J. Cosmol. Astropart.} Phys. 11,
014 (2003).\\
A.-M. M. Abdel Rahman, Gen. Rel. Gravit., 22, 655 (1990).\\
A. I. Arbab, \emph{Class. Quantm. Gravit.} 20, 93 (2003).\\
R. Caldwell, M. Kamionkowski and N. Weinberg, \emph{Phys. Rev.
Lett}. 91, 071301 (2003). \\
R. Caldwell, \emph{Phys.Lett.B}545, 23 (2002).\\
C. Wetterich, \emph{Astron. Astrophys.} 301, 321 (1995).\\
R. R. Caldwell, R. Dave and P. J. Steinhardt, \emph{Phys. Rev.
Lett.} 80, 1582 (1998).\\
P. J. E. Peebles and B. Ratra, \emph{Rev. Mod. Phys.} 75, 559 (2003).\\
P. J. E. Peebles and B. Ratra, \emph{Astrophys. J.} 325, L17 (1988).\\
M. S. Turner and  M. White,  \emph{Phys. Rev. D}, 56, R4439 (1997).\\
S. Perlmutter, \emph{et al.}, \emph{ApJ}, 517, 565 (1998).\\
A. Reiss, \emph{et al.}, \emph{ApJ}, 116, 1009 (1998).\\
 H. Stefancic, \emph{Phys.Lett}. B 586, 5 (2004).\\
 M. P. Dabrowski, T. Stachowiak and M. Szydlowski, \emph{Phys. Rev.D} 68, 103519 (2003).\\
 A. I. Arbab, \emph{Gen. Relt. Gravit.}, 29, 61 (1997).\\
 M. S.Carroll, M. Hoffman and M. Trodden,  \emph{Phys. Rev. D}, 68, 023509
 (2003).\\
  V. Sahni, \emph{Lect.Notes Phys}, 653, 141 (2004).\\
 B. Ratra  and P. J. E. Peebles,   \emph{Phys. Rev.} D 37, 3406 (1988)\\
M. Amarzguioui, Ø. Elgarøy1,  D. F. Mota1 and T. Multamäki,
\emph{A \& A} 454, 707(2006).\\
R. Cai and A. Wang, \emph{Phys. Rev.}D73, 063005
(2006).\\
E. Babichev, V. Dokuchaev, and Yu. Eroshenko, \emph{Phys. Rev.
Lett.} 93, 021102 (2004).\\
\end{document}